\documentclass[aps,prb,reprint,superscriptaddress]{revtex4-1}

\usepackage[utf8]{inputenc}
\usepackage{amsmath,amsthm,amsfonts,amssymb}
\usepackage{leftidx}
\usepackage{bbm}
\usepackage[vcentermath]{youngtab}
\usepackage[colorlinks = true,
linkcolor = blue,
urlcolor  = blue,
citecolor = blue,
anchorcolor = blue]{hyperref}
\usepackage[caption=false]{subfig}
\usepackage{comment}
\usepackage{graphicx}
\usepackage{diagbox}
\usepackage{tikz}
\usetikzlibrary{positioning,scopes,calc,patterns,backgrounds}
\pgfdeclarelayer{myback}
\pgfsetlayers{background,myback,main} 
\tikzset{
  ndot/.style={draw, circle, inner sep=1.2pt},
  dot/.style={draw, circle, fill, inner sep=1.2pt},
  mytext/.style={text width=4cm},
  pattern1/.style={preaction={fill=blue}, pattern=red hatch, pattern color=red, hatch distance=7pt, hatch thickness=2pt},
  pattern2/.style={preaction={fill=green}, pattern=yellow hatch, pattern color=yellow, hatch distance=7pt, hatch thickness=2pt},
  pattern3/.style={preaction={fill=purple}, pattern=orange hatch, pattern color=orange, hatch distance=7pt, hatch thickness=2pt},
  hatch distance/.store in=\hatchdistance,
  hatch distance=5pt,
  hatch thickness/.store in=\hatchthickness,
  hatch thickness=5pt
}
\makeatletter
\pgfdeclarepatternformonly[\hatchdistance,\hatchthickness]{red hatch}
{\pgfqpoint{-1pt}{-1pt}}
{\pgfqpoint{\hatchdistance}{\hatchdistance}}
{\pgfpoint{\hatchdistance-1pt}{\hatchdistance-1pt}}%
{
  \pgfsetcolor{\tikz@pattern@color}
  \pgfsetlinewidth{\hatchthickness}
  \pgfpathmoveto{\pgfqpoint{0pt}{0pt}}
  \pgfpathlineto{\pgfqpoint{\hatchdistance}{\hatchdistance}}
  \pgfusepath{stroke}
}
\pgfdeclarepatternformonly[\hatchdistance,\hatchthickness]{yellow hatch}
{\pgfqpoint{-1pt}{-1pt}}
{\pgfqpoint{\hatchdistance}{\hatchdistance}}
{\pgfpoint{\hatchdistance-1pt}{\hatchdistance-1pt}}%
{
  \pgfsetcolor{\tikz@pattern@color}
  \pgfsetlinewidth{\hatchthickness}
  \pgfpathmoveto{\pgfqpoint{0pt}{0pt}}
  \pgfpathlineto{\pgfqpoint{\hatchdistance}{\hatchdistance}}
  \pgfusepath{stroke}
}
\pgfdeclarepatternformonly[\hatchdistance,\hatchthickness]{orange hatch}
{\pgfqpoint{-1pt}{-1pt}}
{\pgfqpoint{\hatchdistance}{\hatchdistance}}
{\pgfpoint{\hatchdistance-1pt}{\hatchdistance-1pt}}%
{
  \pgfsetcolor{\tikz@pattern@color}
  \pgfsetlinewidth{\hatchthickness}
  \pgfpathmoveto{\pgfqpoint{0pt}{0pt}}
  \pgfpathlineto{\pgfqpoint{\hatchdistance}{\hatchdistance}}
  \pgfusepath{stroke}
}
\makeatother

\newcommand*{\bb}{b}
\newcommand*{\ba}{\tilde{b}}
\newcommand*{\fa}{f}
\newcommand*{\A}{A}
\newcommand*{\B}{B}
\newcommand*{\C}{C}
\newcommand*{\D}{D}
\newcommand*{\SU}[1]{$\mathrm{SU}\hspace{-1mm}\left(#1\right)$}
\DeclareMathOperator{\sgn}{sgn}
\newcommand{\ket}[1]{\left| #1 \right>} 
\newcommand{\abs}[1]{\left| #1 \right|} 

\definecolor{KPnotecolor}{rgb}{0.5,0.,0.5}

\definecolor{FKnotecolor}{rgb}{0.1,0.5,0.8}

\begin{document}


\title{Linear Flavor-Wave Theory for Fully Antisymmetric \SU{N} Irreducible Representations}

\author{Francisco H. Kim}
\affiliation{Institute of Physics, \'Ecole Polytechnique F\'ed\'erale de Lausanne (EPFL), CH-1015 Lausanne, Switzerland}
\author{Karlo Penc}
\affiliation{Institute for Solid State Physics and Optics, Wigner Research Centre for Physics, Hungarian Academy of Sciences, H-1525 Budapest, P.O.B. 49, Hungary}
\affiliation{MTA-BME Lendület Magneto-optical Spectroscopy Research Group, 1111 Budapest, Hungary}
\author{Pierre Nataf}
\affiliation{Institute of Physics, \'Ecole Polytechnique F\'ed\'erale de Lausanne (EPFL), CH-1015 Lausanne, Switzerland}
\author{Fr\'ed\'eric Mila}
\affiliation{Institute of Physics, \'Ecole Polytechnique F\'ed\'erale de Lausanne (EPFL), CH-1015 Lausanne, Switzerland}
\date{\today}
\begin{abstract}
  The extension of the linear flavor-wave theory (LFWT) to fully antisymmetric irreducible representations (irreps) of
  \SU{N} is presented in order to investigate the color order of \SU{N} antiferromagnetic Heisenberg models in several
  two-dimensional geometries. The square, triangular and honeycomb lattices are considered with $m$ fermionic particles
  per site. We present two different methods: the first method is the generalization of the multiboson spin-wave
  approach to \SU{N} which consists of associating a Schwinger boson to each state on a site. The second method adopts
  the Read and Sachdev bosons which are an extension of the Schwinger bosons that introduces one boson for each color and
  each line of the Young tableau. The two methods yield the same dispersing modes, a good indication that they properly
  capture the semi-classical fluctuations, but the first one leads to spurious flat modes of finite frequency not
  present in the second one. Both methods lead to the same physical conclusions otherwise: long-range Néel-type order
  is likely for the square lattice for \SU{4} with two particles per site, but
  quantum fluctuations probably destroy order for more than two particles per site, with $N=2m$. By contrast, 
  quantum fluctuations always lead to corrections larger than the classical order parameter for 
  the tripartite triangular lattice (with $N=3m$) or the bipartite honeycomb lattice (with $N=2m$)
 for more than one particle per site, $m>1$, making the presence of color very unlikely except maybe for $m=2$ on the honeycomb lattice,
 for which the correction is only marginally larger than the classical order parameter.
\end{abstract}


\maketitle

\section{Introduction}

The experimental research with ultra-cold atomic gases in optical lattices is currently a very active and rapidly
progressing field. This type of experiments offers the possibility of fully controlling many parameters, allowing the
realization of a vast number of lattice models at low-temperature. It is thus an important tool to help understand the
many-body physics of quantum nature. In addition to the well-studied systems with \SU{2} symmetry, recent experiments
demonstrate that systems characterized by \SU{N} with $N \leq 10$ can be implemented with up to two particles per site
$m \leq 2$ thanks to the strong decoupling between the electronic angular momentum and the nuclear spin of
alkaline-earth atoms.\cite{gorshkov_two-orbital_2010,scazza_observation_2014,zhang_spectroscopic_2014} The high symmetry
of \SU{N} offers many exciting prospects, such as simulating non-Abelian lattice gauge theories well-known in
high-energy physics or implementing quantum computing schemes. Another aspect of interest is the abundance of exotic
phases that \SU{N} spin Hamiltonian can accommodate.

A simple model that describes the above experimental realization is the fermionic \SU{N} Hubbard model
\begin{equation}
  \label{eq:Hubbard}
  \mathcal{H} = -t \sum\limits_{\left\langle i,j \right\rangle,\mu} \left( c ^{\dagger}_{i,\mu} c ^{}_{j,\mu} + \text{H.c.}\right) + U
  \sum\limits_{i,\mu < \nu} n_{i,\mu} n_{i,\nu},
\end{equation}
where $c ^{\dagger}_{i,\mu}$, $c ^{}_{i,\mu}$ are fermionic operators with $N$ flavors $\mu$ acting on site $i$, thus
generalizing the conventional two flavor spin Hubbard model to $N$ flavors. In the Mott-insulating phase $t \ll U$ with
one particle per site ($m=1$), we obtain the \SU{N} antiferromagnetic (AFM) Heisenberg model
\begin{equation}
  \label{eq:H}
  \mathcal{H} = J \sum\limits_{<i,j>} \sum\limits_{\mu,\nu} \hat{S} ^{\mu}_{\nu}(i) \hat{S} ^{\nu}_{\mu}(j),
\end{equation}
and the operators $\hat{S}^{\mu}_{\nu}$ admit a fermionic representation,
\begin{equation}
  \label{eq:fermionic-rep}
  \hat{S}^{\mu}_{\nu} = \fa^{\dagger}_{\nu} \fa^{}_{\mu} - \frac{m}{N}\delta ^{\mu}_{\nu}.
\end{equation}

This model has been studied in various settings. A Bethe ansatz solution is known in one dimension for any
$N$,\cite{sutherland_model_1975} along with quantum Monte Carlo (QMC) simulation
results.\cite{frischmuth_thermodynamics_1999,messio_entropy_2012,bonnes_adiabatic_2012}
The investigation of higher dimensional systems often relies on many different numerical techniques. The exact
diagonalization\cite{nataf_exact_2016,nataf_exact_2014} can be used for finite cluster sizes, whereas Quantum Monte
Carlo methods\cite{assaad_phase_2005,beach_textsun_2009,cai_quantum_2013,lang_dimerized_2013,wang_competing_2014} can be
applied to problems that do not suffer from the sign problem. The variational Monte Carlo\cite{paramekanti_su_2007,wang_textz_2_2009,lajko_tetramerization_2013,dufour_variational_2015,dufour_stabilization_2016}
and tensor network algorithms\cite{corboz_simultaneous_2011,corboz_spin-orbital_2012,corboz_simplex_2012} have also been
employed for \SU{N} systems, yielding remarkably accurate results. Analytical investigations have also been carried out,
notably using field-theoretical methods in the large-$N$ limit.\cite{read_features_1989}
In particular, chiral spin liquid and valence cluster states are predicted for large $N$ depending on the ratio\cite{hermele_topological_2011,hermele_mott_2009}
\begin{equation}
  k=\frac{N}{m} \,.
\end{equation}
For small values of $N$, however, it was shown using the linear flavor-wave theory (LFWT) and different numerical
methods that the antiferromagnetically ordered phase is
stabilized\cite{toth_three-sublattice_2010,corboz_simultaneous_2011,nataf_exact_2014,luo_spin_2016} for $m=1$, in which
two different colors occupy the adjacent sites of each bond, similar to the spin-$\frac{1}{2}$ Heisenberg square lattice
in a Néel configuration. The LFWT, which originates from the pioneering works of
Papanicolaou\cite{papanicolaou_pseudospin_1984,papanicolaou_unusual_1988} and which was further developed by Joshi
\emph{et al.}\cite{joshi_elementary_1999} and Chubukov\cite{chubukov_fluctuations_1990}, assesses the possibility of a
system to retain a long-range order with quantum fluctuations, and it predicts a magnetic order for $m=1$ up to $N=5$
for the square lattice, and for $N=3$ for the triangular lattice.\cite{tsunetsugu_spin_2006,lauchli_quadrupolar_2006} It
is expected that the magnetic order would be destroyed as $k$ becomes large due to the increase of quantum fluctuations
and the frustration in the system that stems from the extensively degenerate ground-state manifold at the mean-field
level for large $N$. So far, the LFWT has been applied uniquely on the systems with one particle per site ($m=1$), and
it is not yet known if the magnetic order would survive in systems with relatively small $k$ and $m$ with more than one
particle per site ($m>1$).

When placing one particle per site, the $N$ degrees of freedom of \SU{N}, called colors or flavors in reference to
elementary particles, lead to the use of the fundamental irreducible representation of \SU{N}, in which the \SU{N}
matrices act on the $N$-dimensional complex-vector space. However, placing many particles per site can be seen as generating new composite particles
(e.g., quarks giving mesons in particle physics), and the action of \SU{N} has to be adapted to the composite particles,
meaning that a different irreducible representation (irrep) of \SU{N} has to be considered. An irreducible
representation of \SU{N} can be depicted by a Young tableau, labeled by the row lengths ($l_1,\dots,l_k$) or,
alternatively, labeled by a $(N-1)$-tuple whose entry is the difference of the length of the adjacent rows
$[l_1-l_2,l_2-l_3,\dots,l_{k}-0]$. The antisymmetry of the states is represented in the vertically stacked boxes,
whereas the symmetry is represented in the horizontally stacked boxes, leading to the constraint that a Young tableau
cannot have more than $N$ rows $(1 \le k \le N)$. Additionally, a row cannot be longer than the row above it
($l_{k} \le \cdots \le l_{1}$).

\begin{figure}[t!]
  \centerline{\includegraphics[width=0.99\linewidth]{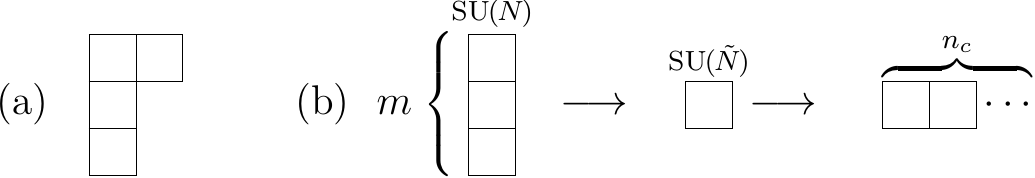} }
  \caption{\label{fig:young-LFWT}(a) A generic Young tableau labeled by [1,0,1] or (2,1,1), a partition of $N=4$. This represents one of the
    irreducible representations of \SU{4}. (b) A Young tableau with $m$ vertical boxes representing the corresponding
    \SU{N} irrep with $\tilde{N}=\binom{N}{m}$ states. These states are mapped to states in
    SU\hspace{-0.5mm}($\tilde{N}$) fundamental irrep, after which the semi-classical approximation $n_{c} \rightarrow \infty$ is performed.}
\end{figure}

In this work, we present two different methods of applying the LFWT to arbitrary irreducible representations, with emphasis on
fully antisymmetric irreps. Such irreps, with a single column of length $m$, are very natural in the context of fermionic cold atoms in optical lattices because 
they describe the Mott phases with $m$ particles per site. Owing to the strong hyperfine interactions, it is
possible to load fermionic atoms with an internal degree of freedom that can take up to $N$ values, thus implementing the \SU{N} symmetry. It is then possible 
to load up to $N$ particles per site, and if the on-site repulsion is strong enough, to stabilize Mott phases with $m$ particles per site for $1 \leq m \leq N$. 
The best candidates are ytterbium, for which $N$ can be as large as 6, and strontium, for which $N$ can be as large as 10.

The first method is an extension of the multiboson
spin-wave~\cite{masashige_transverse_2010,romhanyi_multiboson_2012,penc_spin-stretching_2012} to \SU{N} irreps, where
each state of a given irrep is represented by a boson. A second approach relies on a different bosonic representation of
the states of a given \SU{N} antisymmetric irrep, used by Read and Sachdev.\cite{read_features_1989} Based on the ordered
nature of the ground-state we are considering, we assume a condensate of multiple colors on each sublattice, enabling
the $c$-number substitution of the condensed bosons in the sprit of Bogoliubov.\cite{bogoliubov1947theory} Both
procedures are applied to all the simplest two-dimensional geometries that can accommodate an antiferromagnetic color
order without frustration. When the classical ground-state manifold is infinitely degenerate as in the \SU{3} AFM
Heisenberg model on the square lattice, the LFWT cannot give an accurate prediction of the color order due to the
infrared divergency stemming from the degenerate classical ground-states, although this degeneracy is expected to be
lifted by quantum fluctuations, thus allowing the system to retain a small color order (see
Ref.~\onlinecite{bauer_three-sublattice_2012}).

Henceforth, we consider the square lattice and the honeycomb lattice in a Néel-like two-sublattice configuration
($n_{\text{sub}}=2$), and the triangular lattice with three sublattices ($n_{\text{sub}}=3$), with $n_{\text{sub}}$
being the number of sublattices required for a frustration-free color-order. For an antiferromagnetic Heisenberg model
with a given $N$, it is then natural to consider $m=\frac{N}{n_{\text{sub}}}$ particles per site. We thus apply the
method to the \SU{4} AFM Heisenberg model with $m=2$ on the bipartite square lattice and on the bipartite honeycomb
lattice, and we continue with the \SU{6} AFM Heisenberg model with $m=2$ on the tripartite triangular lattice. We then
derive results for any $N$ on these geometries. We show that $N=4$ on the bipartite square lattice is the only case that
can possess long-range order, in other cases the zero-point quantum fluctuations will destroy the order.

\section{\label{sec:multisun}Multiboson LFWT approach}

We hereby address fully antisymmetric states expressed in terms of $m$ fermions per site. The Young tableau of the
corresponding irrep then consists of a single column with $m$ boxes. In the fundamental representation, the
fermionic representation Eq.~\eqref{eq:fermionic-rep} allows us to write the Heisenberg Hamiltonian, Eq.~\eqref{eq:H},
as
\begin{equation}
  \label{eq:H-fermionic} \mathcal{H} = J \sum\limits_{\langle i,j \rangle} \sum\limits_{\mu,\nu=1}^{N}
\fa^{\dagger}_{\nu,i} \fa^{}_{\mu,i} \fa^{\dagger}_{\mu,j} \fa^{}_{\nu,j},
\end{equation} where the constant term $-\frac{m^{2}}{N}$ has been dropped. The Hamiltonian is then simply a permutation of
two colors from two neighboring sites. $m$ fermionic particles in an antisymmetric configuration form
\begin{equation} \tilde{N}:=\binom{N}{m}
\end{equation} states on a site that transform into themselves according to the corresponding irrep. We can thus assign
a boson to each state of the irrep, providing $\tilde N$ bosons, and we can rewrite the action of the Hamiltonian
(i.e., the color permutation) in the basis of this irrep. This amounts to mapping our original states to
SU\hspace{-0.5mm}($\tilde{N}$) states in the fundamental irrep. The boson that represents the classical ground state can
then be condensed in order to perform the semi-classical expansion by letting $n_{c} \rightarrow \infty$ (see
Fig.~\ref{fig:young-LFWT}). This is analog to the spin-wave expansion where we let $S \rightarrow \infty$. The value of
$n_{c}$ will be set to $1$ at the end of the calculations.

\subsection{\label{sec:su4-m2-sq}\SU{4} $m=2$ on the square lattice}
\begin{figure*}[t!]
  \centerline{\includegraphics[width=0.94\linewidth]{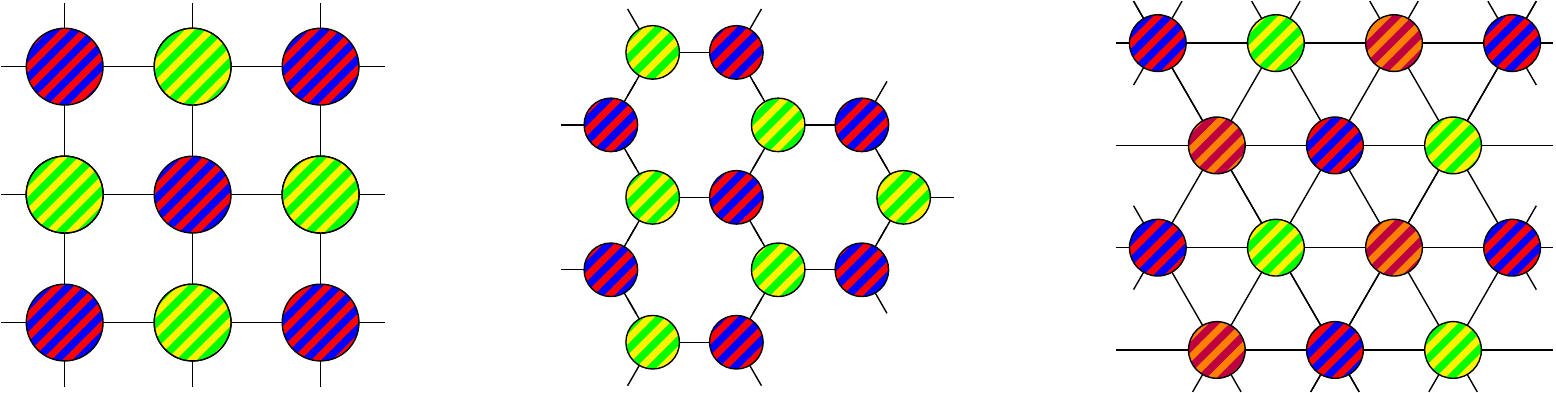} }
  \caption{\label{fig:neel}A Néel-like two-sublattice order on the square lattice and the honeycomb lattice with two particles per site for the \SU{4} AFM
    Heisenberg model, and a three-sublattice order on the triangular lattice with two particles per site for \SU{6}.}
\end{figure*}

Let us first consider \SU{4} with $m=2$ on a bipartite square lattice, on which we assume a Néel-like two-sublattice ordered
state~(see Fig.~\ref{fig:neel}). Furthermore, we assume that the first two colors $A$ and $B$ are condensed on sublattice $\Gamma_{AB}$ and the
last two colors on sublattice $\Gamma_{CD}$. The irrep we are considering is thus $[0,1,0]$. Let the basis of
this six-dimensional irrep be
\begin{align}
  \label{eq:su4-basis}
  &\left\{ \frac{1}{\sqrt{2}}\left( \ket{AB}-\ket{BA} \right), \frac{1}{\sqrt{2}}\left( \ket{AC}-\ket{CA} \right),\right.\nonumber\\
  &\left. \frac{1}{\sqrt{2}}\left( \ket{DA}-\ket{AD} \right), \frac{1}{\sqrt{2}}\left( \ket{BC}-\ket{CB} \right),\right.\nonumber\\
  &\left. \frac{1}{\sqrt{2}}\left( \ket{BD}-\ket{DB} \right), \frac{1}{\sqrt{2}}\left( \ket{CD}-\ket{DC} \right) \right\},
\end{align}
which we label conveniently as the elements of the set 
\begin{equation}
 S:=\{AB,AC,DA,BC,BD,CD\}.
\end{equation}
Note that a different choice of basis does not affect the spectra of the Hamiltonian at the end of the calculations. It
is also worthwhile noting that an orthogonal basis can be systematically found for any irrep of \SU{N} by using the
orthogonal units developed by Young.~\cite{nataf_exact_2014} This yields six Schwinger bosons
$\bb^{\dagger}_{AB}, \bb^{\dagger}_{AC}, \bb^{\dagger}_{DA}, \bb^{\dagger}_{BC}, \bb^{\dagger}_{BD}, \bb^{\dagger}_{CD}$
and their Hermitian counterparts in this basis of the irrep, and they describe the composite particles composed of two
\SU{4} color particles.  The generators for a given site $i$ can be written as
\begin{equation}
  \label{eq:S-irrep}
  \hat{S} ^{\mu}_{\nu}(i) = \sum\limits_{\substack{\alpha=A\\\alpha \neq \mu,\nu}}^{D} \bb^{\dagger}_{\alpha\nu}(i) \bb^{}_{\alpha\mu}(i),
\end{equation}
in which the bosons are antisymmetric in their indices, $\bb^{\dagger}_{\nu\mu}=-\bb^{\dagger}_{\mu\nu}$, such that the indices can be
ordered to yield the aforementioned labels $\{AB,AC,DA,BC,BD,CD\}$. The sign of the permutations takes into account the antisymmetry of the resulting states. As an
example, the operator $\hat{S}^A_C$ is given as
\begin{equation}
  \hat{S} ^{A}_C(i) = -\bb^{\dagger}_{BC}(i) \bb^{}_{AB}(i) -\bb^{\dagger}_{CD}(i) \bb^{}_{DA}(i),
\end{equation}
which is exchanging color $A$ with color $C$ in all the states that allow this transition. The diagonal operator $\hat{S}^C_{C}$
would be given as
\begin{equation}
  \hat{S} ^{C}_C(i) = \bb^{\dagger}_{AC}(i) \bb^{}_{AC}(i) +\bb^{\dagger}_{BC}(i) \bb^{}_{BC}(i) +\bb^{\dagger}_{CD}(i) \bb^{}_{CD}(i).
\end{equation}

This representation of the \SU{N} generators $\hat{S}^{\mu}_{\nu}$ satisfies the \SU{N} commutation relation
\begin{equation}
  \label{eq:sun-comm}
  \left[ \hat{S}^{\alpha}_{\beta}, \hat{S}^{\mu}_{\nu} \right] = \delta ^{\alpha}_{\nu} \hat{S}^{\mu}_{\beta} - \delta ^{\mu}_{\beta} \hat{S}^{\alpha}_{\nu}.
\end{equation}

The Hamiltonian~\eqref{eq:H} can then be written in terms of the Schwinger bosons. This result is obtained by
writing the Hamiltonian Eq.~\eqref{eq:H-fermionic} in the basis of the two-site Hilbert space.

In the language of the composite particles, the constraint $m=2$ can be written as
\begin{equation}
  \label{eq:Mconstraint}
  \sum\limits_{\eta} \bb^{\dagger}_{\eta}(i) \bb^{}_{\eta}(i) = n_{c},
\end{equation}
where $n_{c}=1$ and $\eta \in S$.

Let the site $i_{AB} \in \Gamma_{AB}$ and the site $i_{CD} \in \Gamma_{CD}$. It is now possible to apply the standard
linear flavor-wave theory as in Ref.~\onlinecite{bauer_three-sublattice_2012}. Similar to the $1/S$ expansion in the
spin-wave theory, the limit $n_{c} \rightarrow \infty$ allows us to write
\begin{align}
  \label{eq:hp}
  \bb^{\lambda \dagger}_{\lambda}(i_{\lambda}) \bb^{\lambda}_{\lambda}(i_{\lambda}) &= n_{c} - \sum\limits_{\eta \in S \setminus \{\lambda\}} \bb^{\lambda \dagger}_{\eta}(i_{\lambda}) \bb^{\lambda}_{\eta}(i_{\lambda})\nonumber\\
  \Longrightarrow\ \bb^{\lambda \dagger}_{\lambda}(i_{\lambda}), \bb^{\lambda}_{\lambda}(i_{\lambda}) &\rightarrow \sqrt{n_{c} - \sum\limits_{\eta \in S \setminus \{\lambda\}} \bb^{\lambda \dagger}_{\eta}(i_{\lambda}) \bb^{\lambda}_{\eta}}(i_{\lambda})
\end{align}
where the superscript $\lambda \in L:=\left\{ AB, CD \right\}$ refers to the corresponding sublattice in the spirit of the
Holstein-Primakoff representation. Expanding the square roots in $1/n_{c}$
gives rise to a decomposition of the Hamiltonian in powers of $\sqrt{n_{c}}$:
\begin{equation}
  \mathcal{H} = \mathcal{H}^{(0)} + \mathcal{H}^{(1)} + \mathcal{H}^{(2)} + \mathcal{O}(n_{c}^{\frac{1}{2}}).
\end{equation}
The term $\mathcal{H}^{(0)} \propto n_{c}{}^{2}$ is the classical energy, whereas $\mathcal{H}^{(1)} \propto n_{c}{}^{\frac{3}{2}}$ is the linear
term that vanishes if we start from a classical ground state. In the following, we truncate the Hamiltonian at the
harmonic order and consider the quadratic term $\mathcal{H}^{(2)}$ only. After the Fourier transform

\begin{equation}
  \label{eq:ft}
  \bb^{\lambda}_{\eta}(i_{\lambda}) = \sqrt{\frac{2}{N}} \sum\limits_{\mathbf{k} \in \text{RBZ}} \bb^{\lambda}_{\eta}(\mathbf{k}) e^{-i \mathbf{k}
    \cdot \mathbf{x_i}},
\end{equation}
the quadratic Hamiltonian is given by
\begin{widetext}
  \begin{equation}
    \begin{aligned}
      \mathcal{H}^{(2)} = J n_{c} & \sum _{\mathbf{k}\in \text{RBZ}} \left[ 8 \bb_{CD}^{AB \dagger }(\mathbf{k})\, \bb^{AB}_{CD}(\mathbf{k}) + 8 \bb^{CD \dagger}_{AB}(\mathbf{k})\, \bb^{CD}_{AB}(\mathbf{k}) \right.\\
      &+4\gamma_{\text{sq}}(\mathbf{k}) \bb_{AC}^{AB \dagger }(\mathbf{k})\, \bb^{CD \dagger}_{BD}(-\mathbf{k}) +4 \gamma _{\text{sq}}(\mathbf{k})
      \bb_{AC}^{AB}(\mathbf{k})\, \bb_{BD}^{CD}(-\mathbf{k}) + 4\bb_{AC}^{AB \dagger }(\mathbf{k})\,\bb_{AC}^{AB}(\mathbf{k}) + 4\bb_{BD}^{CD \dagger }(\mathbf{-k})\,\bb_{BD}^{CD}(\mathbf{-k})\\
      &+4\gamma _{\text{sq}}(\mathbf{k}) \bb_{BD}^{AB \dagger}(\mathbf{k})\, \bb_{AC}^{CD \dagger }(-\mathbf{k}) +4\gamma _{\text{sq}}(\mathbf{k})\bb_{BD}^{AB}(\mathbf{k})\,
      \bb_{AC}^{CD}(-\mathbf{k}) + 4\bb_{BD}^{AB \dagger }(\mathbf{k})\,\bb_{BD}^{AB}(\mathbf{k}) + 4\bb_{AC}^{CD \dagger }(-\mathbf{k})\,\bb_{AC}^{CD}(-\mathbf{k})\\
      &+4\gamma _{\text{sq}}(\mathbf{k}) \bb_{DA}^{AB \dagger }(\mathbf{k})\,\bb_{BC}^{CD \dagger}(-\mathbf{k}) +4\gamma _{\text{sq}}(\mathbf{k}) \bb_{DA}^{AB}(\mathbf{k})\,
      \bb_{BC}^{CD}(-\mathbf{k}) + 4\bb_{DA}^{AB \dagger }(\mathbf{k})\,\bb_{DA}^{AB}(\mathbf{k})+ 4\bb_{BC}^{CD \dagger }(-\mathbf{k})\,\bb_{BC}^{CD}(-\mathbf{k})\\
      &\left. +4\gamma _{\text{sq}}(\mathbf{k}) \bb_{BC}^{AB \dagger }(\mathbf{k})\,\bb_{DA}^{CD \dagger}(-\mathbf{k}) +4\gamma _{\text{sq}}(\mathbf{k}) \bb_{BC}^{AB}(\mathbf{k})\,\bb_{DA}^{CD}(-\mathbf{k}) + 4\bb_{BC}^{AB \dagger }(\mathbf{k})\,
        \bb_{BC}^{AB}(\mathbf{k}) + 4\bb_{DA}^{CD \dagger }(-\mathbf{k})\,\bb_{DA}^{CD}(-\mathbf{k}) \right],
    \end{aligned}
  \end{equation}
\end{widetext}
with 
\begin{equation}
  \gamma_{\text{sq}}(\mathbf{k})=\frac{1}{2}(\cos k_x + \cos k_y),
  \label{eq:gammaksquare}
\end{equation}
After the diagonalization of the non-diagonal terms (the only diagonal terms being those with $\bb^{AB}_{CD}$ and
$\bb^{CD}_{AB}$) with the help of an adequate Bogoliubov transformation,
\begin{equation}
  \label{mb-bog}
  \begin{pmatrix}
    \ba ^{AB \dagger}_{AC,\mathbf{k}}\\
    \ba ^{CD }_{BD,-\mathbf{k}}\\
  \end{pmatrix}
  =
  \begin{pmatrix}
    u_{\mathbf{k}} & v_{\mathbf{k}}\\
    v_{\mathbf{k}} & u_{\mathbf{k}}\\
  \end{pmatrix}
  \begin{pmatrix}
    \bb ^{AB \dagger}_{AC,\mathbf{k}}\\
    \bb ^{CD }_{BD,-\mathbf{k}}\\
  \end{pmatrix}
\end{equation}
and similarly for other bosons, the resulting harmonic Hamiltonian reads as
\begin{align}
    \label{eq:H-RBZ}
    \mathcal{H}^{(2)} =& J n_{c} \sum\limits_{\mathbf{k} \in \text{RBZ}} \left[ 8 \left( \bb^{AB \dagger}_{CD}(\mathbf{k}) \bb^{AB}_{CD}(\mathbf{k}) +
        \bb^{CD \dagger}_{AB}(\mathbf{k}) \bb^{CD}_{AB}(\mathbf{k}) \right)\right.\nonumber\\
    &+ \left. \omega_{\text{sq}} (\mathbf{k}) \sum\limits_{\substack{\lambda \in L }} \sum\limits_{\substack{ \eta \in S \setminus L}} \left( \ba^{\lambda
          \dagger}_{\eta}(\mathbf{k}) \ba^{\lambda}_{\eta}(\mathbf{k}) + \frac{1}{2}  \right) \right]\nonumber\\
    &- 16 Jn_{c} \frac{N}{2},
\end{align}
with 
\begin{equation}
  \omega_{\text{sq}}(\mathbf{k}) = 4\sqrt{1 - \abs{\gamma_{\text{sq}}(\mathbf{k})}^{2}}\;.
\end{equation}
The dispersion relation is depicted in Fig.~\ref{fig:dispersion}. Note that it is identical to the dispersion relation of \SU{2}.

Alternatively, in the structural Brillouin zone, we obtain
\begin{equation}
  \begin{aligned}
    \label{eq:H-sq-su4}
    \mathcal{H}^{(2)} &= J n_{c} \sum\limits_{\mathbf{k}} \left\{ 8 \bb^{\dagger}(\mathbf{k})
      \bb^{}(\mathbf{k})\right.\\
    &+ \left. \omega_{\text{sq}} (\mathbf{k}) \sum\limits_{\beta=1}^{4} \left( \ba^{\dagger}_{\beta}(\mathbf{k}) \ba^{}_{\beta}(\mathbf{k}) + \frac{1}{2}  \right) \right\} - 8 Jn_{c}N,
  \end{aligned}
\end{equation}
in which the boson $\bb^{\dagger}$ originates from the decoupled bosons $\bb^{AB}_{CD}$ and $\bb^{AB}_{CD}$, whereas the
bosons $\ba^{\dagger}$ stem from the Bogoliubov transformation in Eq.~\eqref{mb-bog}. We obtain 10 bands in the reduced Brillouin zone, which correspond to 5 bands in the structural Brillouin zone. From the
5 branches, four are dispersive and one is flat at energy $8Jn_{c}$. The (degenerate) dispersive bands are associated to
the possible flavour-exchanges (e.g., $A \leftrightarrow C$ and $A \leftrightarrow D$). The flat band, however,
originates from having the same colors $AB$ (or $CD$) on two neighboring sites of a bond. This is a higher-order
transition, as two colors are different with respect to the chosen classical ground-state. In other words, it requires
the action of two ladder operators: this can be seen in the weight diagram of this irrep, Fig.~\ref{fig:su4-11},
where $AB$ and $CD$ are two edges apart. Thus this higher-order excitation does not interact in the harmonic order of
the bilinear Heisenberg exchange Hamiltonian,\cite{romhanyi_multiboson_2012} and this results in a localized flat band.

The energy per site of the system due to quantum fluctuations is
\begin{align}
  \label{eq:EN}
  E/N &= J n_{c} \left( -8 + 4 \cdot \left\langle \frac{\omega_{\text{sq}}(\mathbf{k})}{2} \right\rangle \right) \nonumber\\
      &= -1.264 J n_{c}\;.
\end{align}

We now define the ordered color moment on the site $i \in \Gamma_{\lambda}$, as
\begin{align}
  \label{eq:ordered-moment}
  m_{i} &= \frac{1}{n_{c}} \left\langle \bb^{\lambda \dagger}_{\lambda}(i) \bb^{\lambda}_{\lambda}(i) \right\rangle\nonumber\\
        &= \frac{1}{n_{c}} \left( n_{c} - \left\langle \sum\limits_{\eta \in S \setminus \{\lambda\}}\bb^{\lambda \dagger}_{\eta}(i) \bb^{\lambda}_{\eta}(i) \right\rangle \right),
\end{align}
so that the fully polarized classical Néel state is $m_i=1$. Then, the reduction of the ordered moment due to quantum
fluctuations is
\begin{align}
  \label{eq:reduction-moment}
  \Delta m_i =& \frac{1}{n_{c}} \left\langle \sum\limits_{\eta \in S \setminus \{\lambda\}} \bb^{\lambda
                \dagger}_{\eta}(i) \bb^{\lambda}_{\eta}(i) \right\rangle\nonumber\\
  =& \left\langle 4 \cdot \frac{1}{2}\left( \frac{4 J n_{c}}{J n_{c} \omega_{\text{sq}}} -1 \right) \right\rangle\nonumber\\
  =&\ 0.786,
\end{align}
where we used the fact that $\left\langle \bb^{AB \dagger}_{CD}(i) \bb^{AB}_{CD}(i) \right\rangle = \left\langle \bb^{CD \dagger}_{AB}(i)
  \bb^{CD}_{AB}(i) \right\rangle = 0$ whereas
$\left\langle \bb^{\lambda \dagger}_{\eta}(i) \bb^{\lambda}_{\eta}(i) \right\rangle$ is finite for
$\lambda \in L$ and $\eta \in S \setminus \{\lambda\}$ as a consequence of the Bogoliubov transformation. This merely reflects the
impossibility for the state $AB$ to fluctuate into the state $CD$ with the bilinear Heisenberg exchange in the harmonic
order.

The ordered moment is then
\begin{align}
  \label{eq:magnetization-sq}
  m_i &= 1 - \Delta m_i = 0.214.
\end{align}
Since the ordered moment $m_{i} > 0$, this theory predicts that the system potentially retains a finite magnetic
order. Note that the correction is not small, however. It is close to 80\%. So, order is likely but not guaranteed.

Note that we could alternatively define the ordered moment as in Ref.~\onlinecite{wang_competing_2014} in which it is
defined on any site $i$ of a bipartite lattice as
\begin{equation}
  \label{eq:ordered-moment-2}
  m_i^{\text{alt}} = \frac{2}{N}\left( \sum\limits_{\mu = 1}^{N/2} S ^{\mu}_{\mu}(i) - \sum\limits_{\mu = \frac{N}{2} + 1}^{N} S ^{\mu}_{\mu}(i) \right),
\end{equation}
giving an ordered moment of $m_i = (-1)^{i}$ for a classical Néel configuration, where the sign depends on the
sublattice. Following this definition, one finds
\begin{equation}
  \label{eq:magnetization-sq-2}
  m_i^{\text{alt}} = (-1)^{i} 0.214.
\end{equation}

\subsection{\label{sec:su4-m2-hex}\SU{4} $m=2$ on the honeycomb lattice}

Following the same construction as in Sec.~\ref{sec:su4-m2-sq}, we assume two sublattices $\Gamma_{AB}$ and $\Gamma_{CD}$, and
$L=\{AB, CD\}$, $S=\{AB,AC,DA,BC,BD,CD\}$ as before~(see Fig.~\ref{fig:neel}). Then the harmonic Hamiltonian for the bipartite honeycomb
lattice can be given as follows:
\begin{equation}
  \label{eq:H-RBZ-hex}
  \begin{aligned}
    &\mathcal{H}^{(2)} = J n_{c} \sum\limits_{\mathbf{k} \in \text{BZ}} \left[ 6 \sum\limits_{\lambda \in L} \sum\limits_{\eta \in L
        \setminus \{\lambda\}} \bb^{\lambda \dagger}_{\eta}(\mathbf{k}) \bb^{\lambda}_{\eta}(\mathbf{k}) \right.\\
    &\left. + \omega_{\text{hon}} (\mathbf{k}) \sum\limits_{\lambda \in L} \sum\limits_{\eta\in S \setminus L} \left( \ba^{\lambda \dagger}_{\eta}(\mathbf{k}) \ba^{\lambda}_{\eta}(\mathbf{k}) + \frac{1}{2}  \right) \right] - 6 Jn_{c}N,
  \end{aligned}
\end{equation}
where $\mathbf{k}$ run over the structural Brillouin zone of the honeycomb lattice, thus giving rise to doubly
degenerate bands. The dispersion relation of the dispersive (``magnetic'') branch (see Fig.~\ref{fig:dispersion}) is given by
\begin{equation}
  \label{eq:w_hex}
  \omega_{\text{hon}}(\mathbf{k}) = 3 \sqrt{1-\abs{\gamma_{\text{hon}}(\mathbf{k})}^{2}},
\end{equation}
where
\begin{equation}
  \label{eq:g-hex}
  \gamma_{\text{hon}}(\mathbf{k})=\frac{1}{3}\left( e^{i k_y} + e^{i\left(\frac{\sqrt{3}}{2}k_x-\frac{1}{2}k_y\right)} + e^{i
      \left(-\frac{\sqrt{3}}{2}k_x-\frac{1}{2}k_y\right)} \right).
\end{equation}

The energy per site due to quantum fluctuations is
\begin{align}
  \label{eq:EN}
  E/N &= J n_{c} \left( -6 + 4 \cdot \left\langle \frac{\omega_{\text{hon}}(\mathbf{k})}{2} \right\rangle\right) \nonumber\\
      &= -1.259 J n_{c}.
\end{align}

The reduction of the ordered moment is
\begin{align}
  \label{eq:moment}
  \Delta m_{i} &= \frac{1}{n_{c}}\left\langle \sum\limits_{\eta=2}^{6} \bb^{1 \dagger}_{\eta}(i) \bb^1_{\eta}(i) \right\rangle\nonumber\\
    &= \frac{1}{n_{c}}\left\langle 4 \cdot \frac{1}{2}\left( \frac{4 J n_{c}}{J n_{c} \omega_{\text{hon}}} -1 \right) \right\rangle \nonumber\\
    &= 1.0328.
\end{align}
The reduction is larger than the classical moment. It is thus likely that no finite order exists on the two-sublattice honeycomb lattice for
$N=4$ with two particles per site.
Note, however, that the reduction is only marginally above 100 \%. So, it cannot be excluded on this basis that a small moment survives quantum fluctuations.

\subsection{\label{sec:su6-m2-trig}\SU{6} $m=2$ on the triangular lattice}
Similar considerations can be done for the triangular lattice for which we assume a three-sublattice order with two
particles per site, i.e., with sublattices $\Gamma_{AB},\Gamma_{CD},\Gamma_{EF}$ where we assume a basis similar to
Eq.~\eqref{eq:su4-basis}~(see Fig.~\ref{fig:neel}).

\begin{figure*}
     \centerline{\includegraphics[width=0.99\linewidth]{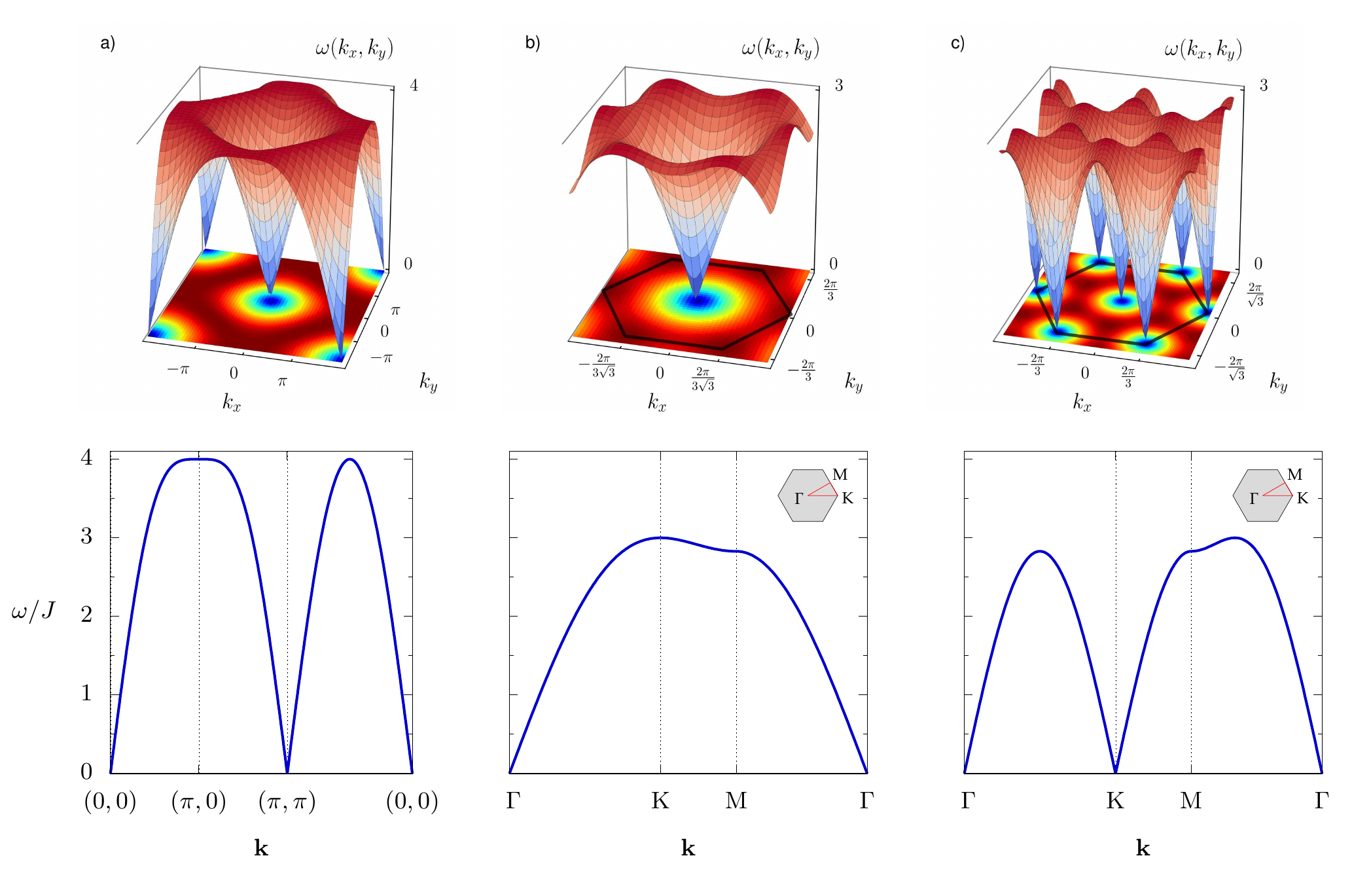} }
  \caption{\label{fig:dispersion}The dispersion relation of the (a) \SU{4} square, (b) \SU{4} honeycomb, and (c) \SU{6} triangular lattice. The first
    structural Brillouin zone is depicted in the 2D heat map. Both the honeycomb and the triangular lattices result in
    the identical first Brillouin zone up to a scaling factor, as the bonds between two given sublattices are identical
    in both lattices up to a scaling factor between the bonds.}
\end{figure*}

Adding all the bonds together and merging them to form bands in the structural Brillouin zone, we obtain
\begin{equation}
  \begin{aligned}
    \label{eq:H-tri}
    \mathcal{H}^{(2)} &= Jn_{c} \sum\limits_{\mathbf{k}} \left[ \sum\limits_{\alpha=1}^{6} 6\bb^{\dagger}_{\alpha}(\mathbf{k}) \bb_{\alpha}(\mathbf{k}) \right.\\
    &+ \left. \omega_{\text{tri}} (\mathbf{k}) \sum\limits_{\beta=1}^{8} \left( \ba^{\dagger}_{\beta}(\mathbf{k})
        \ba^{}_{\beta}(\mathbf{k}) + \frac{1}{2}  \right) \right] - 12 Jn_{c},
  \end{aligned}
\end{equation}
with the dispersion relation~(see Fig.~\ref{fig:dispersion})
\begin{equation}
  \label{eq:g-tri}
  \begin{aligned}
    \omega_{\text{tri}}(\mathbf{k})=3\sqrt{1-\abs{\gamma_{\text{tri}}(\mathbf{k})}^{2}}
  \end{aligned}
\end{equation}

in which
\begin{equation}
  \label{eq:g-tri}
  \begin{aligned}
    \gamma_{\text{tri}}(\mathbf{k}) = \frac{1}{3}\left(e^{i k_{x}} + 2e ^{-i \frac{1}{2} k_x}\cos \frac{\sqrt{3}}{2}k_{y}\right). 
  \end{aligned}
\end{equation}
It is worth noting the similarity between Eqs.~\eqref{eq:g-tri} and \eqref{eq:g-hex}, as the geometric bonds between
two sublattices have the same angle in both the triangular and the honeycomb lattices. We obtain six bands that sit high in energy and eight
bands associated to the exchange of flavors that always keep one flavor on the site, e.g., $AB$ to $AC$.

The energy per site due to quantum fluctuations is
\begin{align}
  \label{eq:EN}
  E/N &= J n_{c} \left( -12 + 8 \cdot \left\langle \frac{\omega_{\text{tri}}(\mathbf{k})}{2} \right\rangle\right) \nonumber\\
      &= -2.518 J n_{c}.
\end{align}

The reduction of the ordered moment is
\begin{align}
  \label{eq:moment}
  \Delta m_{i} &= \frac{1}{n_{c}} \left\langle \sum\limits_{\eta=2}^{15} \bb^{1 \dagger}_{\mu}(i) \bb^1_{\mu}(i) \right\rangle\nonumber\\
    &= \frac{1}{n_{c}} \left\langle 8 \cdot \frac{1}{2}\left( \frac{3 J n_{c}}{J n_{c} \omega_{\text{tri}}(\mathbf{k})} -1 \right) \right\rangle \nonumber\\
    &= 2.066,
\end{align}
hence, we can conclude that the long-range color order is almost certainly destroyed by quantum fluctuations.

\subsection{General $m$}
\label{sec:general-m}

For any antisymmetric \SU{N} irrep with $m$ particles, the generators $\hat{S} ^{\mu}_{\nu}$ for a given site $i$ can be written as
\begin{align}
  \hat{S}^{\mu}_{\nu}(i) &= \sum\limits_{\substack{\alpha_1,\dots,\alpha_m\\\alpha_1,\dots,\alpha_m \neq \mu,\nu}}
  \sgn(\sigma_1)\sgn(\sigma_2) \nonumber\\
                         &\times \bb^{\dagger}_{\sigma_1\cdot(\alpha_1 \dots \alpha_m \nu)}(i) \bb^{}_{\sigma_2\cdot(\alpha_1 \dots \alpha_m \mu)}(i),
\end{align}
with $\alpha_1, \dots, \alpha_m$ run over the $N$ colors and $\sigma_{1},\sigma_{2}$ are permutations that order the
letters in the alphabetical order. This is a direct generalization of Eq.~\eqref{eq:S-irrep}, and its action is the
permutation of color $\mu$ with $\nu$ while taking care of the sign change due to the antisymmetry of the states. Note
that this can be generalized further for any general irrep by determining the action of the generator
$\hat{S}^{\mu}_{\nu}$ on the basis states of the irrep.

From the three models above, we observe the existence of dispersive branches and non-zero flat branches at the harmonic
level of the Hamiltonian. The dispersive branches stem from the transitions occurring from exactly one color exchange
between two neighboring sites. In the case of the bipartite \SU{4} square lattice, the state $AB$ can decay into four
different states ($AC,DA,BC,BD$) when exchanging one color with the neighboring state $CD$, yielding four dispersive
branches. However, going from $AB$ to $CD$ requires the exchange of two colors at least, resulting in a flat band in the
harmonic order with an energy sitting at $2Jz$, i.e., the energy cost of exchanging two colors with $z$ possible nearest
neighbors.

In general, we can have bands with energy $nzJn_{c}$ ($n \in \{2,\dots,m\}$) depending on the number of the required
color exchanges for a possible target state. Consequently, it is possible to deduce the diagonalized quadratic
Hamiltonian by determining the number of color exchanges that are needed for every possible transition. In general, for
any $m$ with $k=2$ for the square or $k=3$ for the triangular lattice, the quadratic
Hamiltonian is given by

\begin{widetext}
  \begin{equation}
    \label{eq:H-gen}
    \mathcal{H}^{(2)} = J n_{c} \sum\limits_{\mathbf{k}} \left\{ \sum\limits_{n=2}^{m} n z
      \sum\limits_{\alpha=1}^{\binom{m}{n} \binom{N-m}{n}} \bb^{\dagger}_{\alpha}(\mathbf{k}) \bb^{}_{\alpha}(\mathbf{k}) + \omega_{\text{sq/tri}}(\mathbf{k}) \sum\limits_{\beta=1}^{m(N-m)} \left( \ba^{\dagger}_{\beta}(\mathbf{k}) \ba^{}_{\beta}(\mathbf{k}) + \frac{1}{2}  \right) \right\} - \frac{m(N-m)}{2}zJn_{c}N,
  \end{equation}
\end{widetext}
where the sum runs over the structural Brillouin zone, and $z$ is the coordination number between two sublattices ($z=4$
for the square lattice and $z=3$ for the triangular lattice). The dimension of the considered antisymmetric irrep
$[m,0,\dots]$ is $\binom{N}{m}$. The use of the Holstein-Primakoff bosons with the limit $n_{c} \rightarrow \infty$
leads to $\binom{N}{m}-1$ branches in the structural Brillouin zone, of which $\binom{m}{1}\binom{N-m}{1}=m(N-m)$
branches are dispersive. Since $N=m n_{\text{sub}}$ for a given value of $m$, the square lattice will have $m^{2}$
dispersive branches and the triangular lattice will have $2m^{2}$ branches. Hence, we can conclude that for a given
number of particles per site $m$, the reduction of the magnetization $\Delta m_{i}$ is given by
\begin{equation}
  \label{eq:reduction-moment-sq}
  \begin{aligned}
    \Delta m_{i}^{\text{sq}}(m) =& m^{2} \left\langle \frac{1}{2}\left( \frac{4 J n_{c}}{J n_{c} \omega_{\text{sq}}} -1 \right) \right\rangle\nonumber\\
    =&\ 0.197 m^{2}
  \end{aligned}
\end{equation}
for the square lattice, and
\begin{equation}
  \label{eq:reduction-moment-tri}
  \begin{aligned}
    \Delta m_{i}^{\text{tri}}(m) &= 2 m^{2} \left\langle \frac{1}{2}\left( \frac{3 J n_{c}}{J n_{c} \omega_{\text{tri}}(\mathbf{k})} -1 \right) \right\rangle \nonumber\\
    &= 0.516 m^{2}
  \end{aligned}
\end{equation}
for the triangular lattice.

As for the flat
modes, there are $\binom{m}{n}\binom{N-m}{n}$ flat branches at energy $nzJn_{c}$, with $n \in \{ 2,\dots,m \}$ being the
number of color-exchange applied at a state.

The same conclusion also applies for the honeycomb lattice, with the only difference being the number of branches that is
doubled in the first structural Brillouin zone. Introducing an index $\xi$ to account for the doubling of the branches, we
obtain
\begin{widetext}
  \begin{equation}
    \label{eq:H-gen}
    \mathcal{H}^{(2)} = J n_{c} \sum\limits_{\mathbf{k}} \sum\limits_{\xi=1}^{2} \left\{ \sum\limits_{n=2}^{m} n z
      \sum\limits_{\alpha=1}^{\binom{m}{n} \binom{N-m}{n}} \bb^{\dagger}_{\alpha,\xi}(\mathbf{k}) \bb^{}_{\alpha,\xi}(\mathbf{k}) + \omega_{\text{hon}} (\mathbf{k}) \sum\limits_{\beta=1}^{m(N-m)} \left( \ba^{\dagger}_{\beta,\xi}(\mathbf{k}) \ba^{}_{\beta,\xi}(\mathbf{k}) + \frac{1}{2}  \right) \right\} - \frac{m(N-m)}{2}zJn_{c}N
  \end{equation}
\end{widetext}
for the honeycomb lattice, where $z=3$. Hence, the reduction of the magnetization as a function of the number of
particles per site is given by
\begin{equation}
  \label{eq:reduction-moment-sq}
  \begin{aligned}
    \Delta m_{i}^{\text{hon}}(m) =& m^{2} \left\langle \frac{1}{2}\left( \frac{4 J n_{c}}{J n_{c} \omega_{\text{hon}}(\mathbf{k})}
        -1 \right) \right\rangle \nonumber\\
    =&\ 0.258 m^{2}.
  \end{aligned}
\end{equation}

In all cases, the reduction of the local order parameter is much larger than 1 for $m\geq 3$, making the presence of 
long-range order very unlikely.


\section{\label{sec:altern-boson-repr}Read and Sachdev bosonic representation}

Harmonic fluctuations can be analyzed with an alternative approach by using a different bosonic representation for the
\SU{N} generators. This bosonic representation briefly mentioned in Ref.~\onlinecite{read_features_1989} is an extension
of the Schwinger bosons, and can be applied to any irreps whose Young tableaux contain $m$ rows and $n_{c}$ columns. It
assumes one boson for each color as well as for each row of the Young tableau, and the bosons are then antisymmetrized
in accordance with the chosen irrep. In this realization, the \SU{N} operators can be written as

\begin{equation}
  \label{eq:RS-op}
  \hat{S}^{\mu}_{\nu} = \sum\limits_{a=1}^{m} \bb_{\nu a}^{\dagger} \bb_{\mu a} - \frac{n_c}{2}\delta_{\mu \nu},
\end{equation}
where $\mu,\nu\in\{A,B,\dots\}\equiv\{1,\dots,N\}$ are the color indices and $a \in \{1,\dots,m\}$ are the row indices. They naturally satisfy the \SU{N} commutation
relations. The constraints
\begin{equation}
  \label{eq:RS-constraint}
  \sum\limits_{\alpha=1}^{N} \bb ^{\dagger}_{\alpha a} \bb ^{}_{\alpha b} = \delta_{ab}n_{c},
\end{equation}
with $a \in \{1,\dots,N\}$ and $a,b \in \{1,\dots,m\}$ ensure that we work in the given irrep. The constraints that involve the same
line indices are the same as the constraints of the Schwinger bosons, whereas the other equations are additional constraints
that enforce the antisymmetry of the irrep.

The Heisenberg Hamiltonian in this bosonic representation is given by
\begin{equation}
  \begin{aligned}
    \label{eq:RS-H}
    \mathcal{H} =& J \sum\limits_{<i,j>} \sum\limits_{\mu,\nu} \hat{S} ^{\mu}_{\nu}(i) \hat{S} ^{\nu}_{\mu}(j)\\
    =& J \sum\limits_{\substack{<i,j> \\ \mu,\nu}} \sum\limits_{a,b=1}^{m} \bb ^{\dagger}_{\nu a}(i) \bb^{}_{\mu a}(i) \bb ^{\dagger}_{\mu b}(j) \bb^{}_{\nu b}(j).
  \end{aligned}
\end{equation}

\subsection{\SU{4} $m=2$ on the square lattice}
\label{sec:RS-square}
We now turn our attention to the square lattice with $m=2$. Let us assume an ordered state in which the colors $\A$ and
$\B$ sit on the sublattice $\Lambda_{AB}$ and the colors $\C$ and $\D$ are on the sublattice $\Lambda_{CD}$ of the
square lattice. Note that we have deliberately broken the symmetry by choosing specific colors for the sublattice. In
the limit $n_c \rightarrow \infty$, we assume that there is a condensate of colors $\A$ and $\B$ on the site $i$ and a
condensate of colors $\C$ and $\D$ on the site $j$. Consequently, it is possible to perform the Bogoliubov substitution
of the condensed bosons with $c$-numbers (with $c\in \mathbb{C}$), i.e.,

\begin{equation}
  \label{eq:RS-condensate}
  \begin{aligned}
    \bb^{\dagger}_{A a}(i) \rightarrow z^{*}_{A a}, \qquad \bb^{\dagger}_{B a}(i) \rightarrow z^{*}_{B a},\\
    \bb^{\dagger}_{C a}(j) \rightarrow z^{*}_{C a}, \qquad \bb^{\dagger}_{D a}(j) \rightarrow z^{*}_{D a},\\
  \end{aligned}
\end{equation}
for any $i\in\Lambda_{AB},\ j\in\Lambda_{CD}$, and $a\in\{1,\dots,m\}$. This replacement is true when
considering the expectation value of the bosonic number operators and the operators $S ^{\mu}_{\nu}$.
It is also worthwhile noting that the conventional \SU{2} spin-wave theory in the
harmonic order also corresponds to replacing the condensed bosons by a $c$-number.

In this limit of the large condensate $n_c \rightarrow \infty$, the constraints~\eqref{eq:RS-constraint} for the sublattice
$\Lambda_{\A\B}$ to order $\mathcal{O}(n_c)$ are reduced to

\begin{equation}
  \label{eq:RS-constraint-su4-m2-i}
  \begin{cases}
    z^{*}_{A1} z^{}_{A1} + z^{*}_{B1} z^{}_{B1} = n_{c}\\
    z^{*}_{A2} z^{}_{A2} + z^{*}_{B2} z^{}_{B2}= n_{c}\\
    z^{*}_{A1} z^{}_{A2} + z^{*}_{B1} z^{}_{B2}= 0.\\
  \end{cases}
\end{equation}
The complex-conjugate counterpart of the third equation in Eq.~\eqref{eq:RS-constraint-su4-m2-i} has been dropped as
they are equivalent.

When written in a matrix form $U^{AB}$ such that
\begin{equation}
  \label{eq:RS-constraint-U-i}
  z_{\mu a} =: \sqrt{n_c} \, [U^{AB}]_{\mu a}
\end{equation}
with $\mu~\in~\{A,B\}$ (the first $\frac{N}{2}$ colors) and $a~\in~\{1,2\}~\equiv~\{1,\dots,m\}$, the set of equations
Eq.~\eqref{eq:RS-constraint-su4-m2-i} amounts to imposing a unitarity condition on the matrix $U^{AB}$. Alternatively,
the matrix elements of this unitary matrix can be parametrized in the following way. The set of equations
Eq.~\eqref{eq:RS-constraint} can be written as
\begin{equation}
  \label{eq:RS-constraint-su4-m2V2}
  \begin{cases}
   \sum_{a,b} z^{*}_{Aa} \delta_{a,b} z^{}_{Ab} + \sum_{a,b} z^{*}_{Ba} \delta_{a,b} z^{}_{Bb}= 2 n_{c},\\
    \sum_{a,b}z^{*}_{Aa} \sigma^{(\alpha)}_{a,b} z^{}_{Ab} + \sum_{a,b} z^{*}_{Ba} \sigma^{(\alpha)}_{a,b} z^{}_{Bb}= 0,  \end{cases}
\end{equation}
where $\sigma^{(\alpha)}_{a,b}$ are Pauli matrices with $\alpha = x,y,z$
or
\begin{equation}
  \label{eq:RS-constraint-su4-m2V3}
  \begin{cases}
    \mathbf{z}^{*}_{A}\cdot \mathbf{z}^{}_{A} + \mathbf{z}^{*}_{B} \cdot \mathbf{z}^{}_{B}= 2 n_{c},\\
        \mathbf{z}^{*}_{A}\cdot \sigma^{(\alpha)} \cdot \mathbf{z}^{}_{A} + \mathbf{z}^{*}_{B} \cdot \sigma^{(\alpha)} \cdot \mathbf{z}^{}_{B}= 0.  \end{cases}
\end{equation}
 We can think of the problem as having two antiferromagnetically alligned \SU{2} spins (the $\A$ and the $\B$), the
 $(z^{*}_{A1},z^{*}_{A2})$ and $(z^{*}_{B1},z^{*}_{B2})$ being the \SU{2} spinors of the two spins, and they can be parametrized as
\begin{align}
  \label{eq:RS-zsubi}
  z_{A1} &= \sqrt{n_c}\, e^{i \chi_{AB}} \cos{\frac{\vartheta_{AB}}{2}},  \nonumber\\ 
  z_{A2} &= \sqrt{n_c}\, e^{i \chi_{AB}} \sin{\frac{\vartheta_{AB}}{2}} e^{-i \varphi_{AB}}, \nonumber\\
  z_{B1} &= \sqrt{n_c}\, \sin{\frac{\vartheta_{AB}}{2}},  \nonumber\\
  z_{B2} &= -\sqrt{n_c}\, \cos{\frac{\vartheta_{AB}}{2}} e^{-i \varphi_{AB}}
\end{align}
when condensed.

The same consideration can be done for the sublattice $\Lambda_{\C\D}$, starting from the constraints
Eq.~\eqref{eq:RS-constraint} in the limit of the large $n_{c}$:
\begin{equation}
  \label{eq:RS-constraint-su4-m2-j}
  \begin{cases}
    z^{*}_{C1} z^{}_{C1} + z^{*}_{D1} z^{}_{D1} = n_{c}\\
    z^{*}_{C2} z^{}_{C2} + z^{*}_{D2} z^{}_{D2}= n_{c}\\
    z^{*}_{C1} z^{}_{C2} + z^{*}_{D1} z^{}_{D2}= 0.\\
  \end{cases}
\end{equation}

This can be rewritten further in a unitary matrix form $U^{CD}$:
\begin{equation}
  \label{eq:RS-constraint-U-j}
  z_{\mu a} =: \sqrt{n_c} \, [U^{CD}]_{\mu a}
\end{equation}
with $\mu~\in~\{C,D\}$ (the last $\frac{N}{2}$ colors) and $a~\in~\{1,2\}~\equiv~\{1,\dots,m\}$, or alternatively, with the following parametrization:
\begin{align}
  \label{eq:RS-zsubj}
  z_{C1} &= \sqrt{n_c}\, e^{i \chi_{CD}} \cos{\frac{\vartheta_{CD}}{2}},  \nonumber\\ 
  z_{C2} &= \sqrt{n_c}\, e^{i \chi_{CD}} \sin{\frac{\vartheta_{CD}}{2}} e^{-i \varphi_{CD}}, \nonumber\\
  z_{D1} &= \sqrt{n_c}\, \sin{\frac{\vartheta_{CD}}{2}},  \nonumber\\
  z_{D2} &= -\sqrt{n_c}\, \cos{\frac{\vartheta_{CD}}{2}} e^{-i \varphi_{CD}} .
\end{align}

Following this procedure, the bosons
$\bb^{(\dagger)}_{\A a}(i),\,\bb^{(\dagger)}_{\B a}(i),\,\bb^{(\dagger)}_{\C a}(j),\,\bb^{(\dagger)}_{\D a}(j)$ can be
finally replaced by their corresponding $c$-numbers in the Hamiltonian Eq.~\eqref{eq:RS-H}, yielding a quadratic
Hamiltonian $\mathcal{H}^{(2)}$ of the order $\mathcal{O}(n_c)$. After Fourier-transforming,
\begin{equation}
  \label{eq:RS-FT}
  \bb^{}_{\mu a}(i) = \sqrt{\frac{2}{N_{\text{sites}}}} \sum\limits_{\mathbf{k}\in \text{RBZ}} \bb^{}_{\mu a}(\mathbf{k})
\end{equation}
with $N_{\text{sites}}$ being the number of sites, the quadratic Hamiltonian suited for the generalized Bogoliubov
transformation is then given by
\begin{align}
  \label{eq:RS-H2-nondiag}
  \mathcal{H}^{(2)} = \frac{zJn_{c}}{2} \sum\limits_{\mathbf{k}\in \text{RBZ}}
  \left( \mathbf{\bb}^{\dagger}_{\mathbf{k}},\ltrans{\mathbf{\bb}^{}_{\mathbf{-k}}} \right)
  M_{\mathbf{k}}
  \begin{pmatrix}
    \mathbf{\bb}^{}_{\mathbf{k}}\\
    \ltrans{\mathbf{\bb}^{\dagger}_{\mathbf{-k}}}
  \end{pmatrix}
  - 2zJn_{c}N,
\end{align}

with $z=4$ the coordination number and
\begin{subequations}
\begin{align}
  \mathbf{\bb}^{\dagger}_{\mathbf{k}} =& \Bigl(\bb^{\dagger}_{\C 1}(\mathbf{k}), \bb^{\dagger}_{\C 2}(\mathbf{k}),
                                         \bb^{\dagger}_{\D 1}(\mathbf{k}), \bb^{\dagger}_{\D 2}(\mathbf{k}), \Bigr. \nonumber\\
 & \qquad \Bigl. \bb ^{\dagger}_{\A 1}(\mathbf{k}), \bb ^{\dagger}_{\A 2}(\mathbf{k}), \bb ^{\dagger}_{\B 1}(\mathbf{k}), \bb ^{\dagger}_{\B 2}(\mathbf{k}) \Bigr),\\
  \mathbf{\bb}_{\mathbf{-k}} =& \ltrans{\Bigl(\bb_{\C 1}(\mathbf{-k}), \bb_{\C 2}(\mathbf{-k}), \bb_{\D 1}(\mathbf{-k}),
                                \bb_{\D 2}(\mathbf{-k}), \Bigr.}\nonumber\\
  & \qquad \Bigl. \bb_{\A 1}(\mathbf{-k}), \bb_{\A 2}(\mathbf{-k}), \bb_{\B 1}(\mathbf{-k}), \bb_{\B 2}(\mathbf{-k}) \Bigr),\\
  M_{\mathbf{k}} =& \frac{1}{2}
                    \begin{pmatrix}
                      \mathbbm{1}_{8} & B_{\mathbf{k}}\\
                      B^{\dagger}_{\mathbf{k}} & \mathbbm{1}_{8}
                    \end{pmatrix},\label{eq:bog-Mdef}\\
  \label{eq:RS-ABtilde}
  B_{\mathbf{k}} =&
                    \begin{pmatrix}
                      0 & \gamma^{*}_{k} U^{\intercal}\\
                      \gamma_{k} U & 0
                    \end{pmatrix}.
\end{align}
\end{subequations}
The geometrical factor $\gamma_{\mathbf{k}}$ is defined in Eq.~(\ref{eq:gammaksquare}) with the property that
$\gamma_{-\mathbf{k}}=\gamma^{*}_{\mathbf{k}}$, and the matrix $U$ stems from $U_{AB}$ and $U_{CD}$:
\begin{equation}
  \label{eq:RS-jointU}
  U =
  \begin{pmatrix}
    z_{A1}z_{C1} & z_{A2}z_{C1} & z_{A1}z_{D1} & z_{A2}z_{D1}\\
    z_{A1}z_{C2} & z_{A2}z_{C2} & z_{A1}z_{D2} & z_{A2}z_{D2}\\
    z_{B1}z_{C1} & z_{B2}z_{C1} & z_{B1}z_{D1} & z_{B2}z_{D1}\\
    z_{B1}z_{C2} & z_{B2}z_{C2} & z_{B1}z_{D2} & z_{B2}z_{D2}\\
  \end{pmatrix}
\end{equation}
i.e., it is equal to $U_{AB} \otimes U^{\intercal}_{CD}$ with permuted columns, and is thus also unitary. Note that the structure of the
matrix $M_{\mathbf{k}}$ above is true in general for any $N$ and corresponding $m$ for any of the three lattices
considered in this work, as this is a consequence of the structure of the Hamiltonian in
Eq.~\eqref{eq:RS-H2-nondiag}.

Using the matrix $Y$,
\begin{equation}
  \label{eq:RS-Y}
  Y=
  \begin{pmatrix}
    \mathbbm{1}_{8} & 0\\
    0 & \mathbbm{-1}_{8}\\
  \end{pmatrix},
\end{equation}
the generalized Bogoliubov transformation reduces to searching the eigenvalues $\lambda \rightarrow \frac{1}{2}\omega_{\mathbf{k}}$ of the matrix
$YM_{\mathbf{k}}$. The eigenvalues can be easily found thanks to the simple block structure of this matrix. With the
identity that
\begin{equation}
  \label{eq:RS-BdagB}
  B_{\mathbf{k}}^{\dagger}B_{\mathbf{k}} =
  \begin{pmatrix}
    \abs{\gamma_{\mathbf{k}}}^{2} \mathbbm{1}_{8} & 0\\
    0 & \abs{\gamma_{\mathbf{k}}}^{2} \mathbbm{1}_{8}\\
  \end{pmatrix}
\end{equation}
for any unitary matrix $U$, it results that
\begin{equation}
  \begin{aligned}
  \label{eq:RS-YMYM}
  YM_{\mathbf{k}}YM_{\mathbf{k}} =& \frac{1}{4}\left(1-\abs{\gamma_{\mathbf{k}}}^{2}\right) \mathbbm{1}_{16}\\
  =& \lambda^{2}\, \mathbbm{1}_{16}.
  \end{aligned}
\end{equation}
The eigenvalues are then given by
\begin{equation}
  \label{eq:RS-lambda}
  \lambda = \pm \frac{1}{2} \sqrt{1-\abs{\gamma_{\mathbf{k}}}^{2}}.
\end{equation}
By compactifying the notation, we finally find the diagonalized quadratic Hamiltonian

\begin{equation}
  \label{eq:RS-H2-diag}
  \mathcal{H}^{(2)} = Jn_{c} \sum\limits_{\mathbf{k}\in \text{RBZ}} \omega(\mathbf{k}) \sum\limits_{\mu=1}^{8} \left(
    \ba^{\dagger}_{\mu}(\mathbf{k}) \ba^{}_{\mu}(\mathbf{k}) + \frac{1}{2} \right)- 2zJn_{c}N,
\end{equation}
in which the dispersion relation is given by

\begin{equation}
  \label{eq:RS-w2}
\omega(\mathbf{k}) = z \sqrt{1 - \abs{\gamma_{\mathbf{k}}}^{2}}.
\end{equation}
This yields the same dispersive branches as in the previous calculations in Sec.~\ref{sec:su4-m2-sq} without the flat branches.

The different choices of the set of parameters $\vartheta_{AB},\varphi_{AB},\chi_{AB}$ or
$\vartheta_{CD},\varphi_{CD},\chi_{CD}$ are all related by unitary transformations, hence they result in a unitary
transformation of the matrix $U$ in Eq.~\eqref{eq:RS-ABtilde}. However, since Eq.~\eqref{eq:RS-BdagB} holds for any
unitary matrix $U$, it follows that any unitary transformation on $U$ leaves the eigenvalues of $M_{\mathbf{k}}$
invariant, i.e., any solution that satisfies the modified constraints Eq.~\eqref{eq:RS-constraint-su4-m2-i} leads to the
same dispersion relation in Eq.~\eqref{eq:RS-w2} after the Bogoliubov transformation. Thus, there exists a gauge degree
of freedom $U(m)$ for each sublattice.

As an example, the solution
\begin{equation}
  \label{eq:RS-condensate-1}
  \begin{aligned}
    \bb^{\dagger}_{A 1}(i), \bb^{\dagger}_{C 1}(j) \rightarrow \sqrt{\frac{n_{c}}{2}},\quad \bb^{\dagger}_{B 1}(i),\bb^{\dagger}_{D 1}(j) \rightarrow \sqrt{\frac{n_{c}}{2}},\\
    \bb^{\dagger}_{A 2}(i),\bb^{\dagger}_{C 2}(j) \rightarrow - \sqrt{\frac{n_{c}}{2}}, \quad \bb^{\dagger}_{B 2}(i),\bb^{\dagger}_{D 2}(j) \rightarrow \sqrt{\frac{n_{c}}{2}},\\
  \end{aligned}
\end{equation}
yields the following matrix $\tilde{B}_{\mathbf{k}}$ in Eq.~\eqref{eq:RS-ABtilde}:
\begin{equation}
  \label{eq:RS-B1}
  \tilde{B}_{\mathbf{k}} = \frac{\gamma_{\mathbf{k}}}{2} 
                            \begin{pmatrix}
                              1 & -1 & 1 & -1\\
                              -1 & 1 & 1 & -1\\
                              1 & 1 & 1 & 1\\
                              -1 & -1 & 1 & 1
  \end{pmatrix},
\end{equation}
which in turn results in the dispersion relation~\eqref{eq:RS-w2} after the Bogoliubov transformation.

\subsection{Arbitrary $m$ on different lattices}
The analysis in Sec.~\ref{sec:RS-square} can be straightforwardly generalized to any $N$ and $m$ for a two-sublattice
order, i.e., on the square or honeycomb lattice. This is also easily applied to the three-sublattice order on the
triangular lattice. The only difference with the two-sublattice order is in the Hamiltonian generated after the
$c$-number replacement of the condensed bosons. Unlike in the two-sublattice order calculations where the resulting
Hamiltonian is purely quadratic, higher-order terms are generated in the Hamiltonian, i.e.,
\begin{equation}
  \label{eq:RS-H-3}
  \mathcal{H} = \mathcal{H}^{(2)} + \mathcal{H}^{(3)} + \mathcal{H}^{(4)},
\end{equation}
where $\mathcal{H}^{(2)} \propto \mathcal{O}(n_{c})$, $\mathcal{H}^{(1)} \propto \mathcal{O}(n_{c}^{\frac{1}{2}})$ and
$\mathcal{H}^{(1)} \propto \mathcal{O}(1)$. However, once we truncate the Hamiltonian to keep only the dominant term of the
order $\mathcal{O}(n_c)$, the rest of the calculations are identical to Sec.~\ref{sec:RS-square}. Hence, the procedure
can be applied to any of the three lattice geometries considered in this work. For given $N$, $m$ and assuming a
color-ordered ground-state on one of the three lattices we considered, let us denote the color index of one of the
condensed colors on each sublattice $l\in\{1,\dots,k\}$ by $\mu^{l}\in\{1,\dots,m\}$, and $a,b\in\{1,\dots,m\}$. In the
limit $n_c \rightarrow \infty$, this allows one to use the Bogoliubov prescription of replacing the bosons by a
$c$-number, provided that the numbers satisfy the antisymmetrization constraints~\eqref{eq:RS-constraint}. In the
large-$n_{c}$ limit, these constraints become
\begin{equation}
  \label{eq:RS-constraint-macroscopic}
  \sum\limits_{\mu^{l}=1}^{m} z^{l*}_{\mu a} z^{l}_{\mu b} = \delta_{ab}n_{c}.
\end{equation}
for each sublattice $i$ with corresponding condensed boson colors $\mu$. One particular solution that satisfies the
constraints Eq.~\eqref{eq:RS-constraint-macroscopic} are
\begin{equation}
  \label{eq:RS-constraint-gen-sol}
  \begin{aligned}
    z^{l}_{\mu a} &\rightarrow \varphi^{l}_{\mu a}(m) \sqrt{\frac{n_c}{m}} := \sqrt{n_c} U^{l}_{\mu a},
  \end{aligned}
\end{equation}
with the phase $\varphi^{l}_{\mu a}(m)$ defined by
\begin{equation}
  \label{eq:RS-constraint-gen-sol}
  \varphi^{l}_{\mu a}(m) := e^{-i (a-1)\frac{2 \pi}{m}\mu}.
\end{equation}
It can be easily verified that $z^{}_{\mu a}$ satisfies the constraints by using the identity
$\sum\limits_{k=0}^{n-1} e^{q\frac{2 \pi i}{n}k}=0$, where $n\in\mathbb{N}_{>2}$ and $q\in\{1,\dots,n-1\}$. An example
of the phases for four condensed bosons per site ($m=4$) for \SU{8} on the square/honeycomb lattice or for \SU{12} on the
triangular lattice is shown in Table~\ref{tab:phase}. Any unitary transformation on the matrix $U^{l}$ yields a solution
of Eq.~\eqref{eq:RS-constraint-macroscopic}.

\begin{table}[!htb]
  \centering
  \begin{ruledtabular}
    \begin{tabular}{|c|cccc|}
      $\mu \in \{A,B,C,D\}$ & $a=1$ & $a=2$ & $a=3$ & $a=4$ \\
      \hline
      $A$ & 0 & \(e^{-i x}\) & \(e^{-2i x}\) & \(e^{-3i x}\)\\
      $B$ & 0 & \(e^{-2ix}\) & \(e^{-4ix}\) & \(e^{-6 ix}\)\\
      $C$ & 0 & \(e^{-3ix}\) & \(e^{-6ix}\) & \(e^{-9 ix}\)\\
      $D$ & 0 & \(e^{-4ix}\) & \(e^{-8ix}\) & \(e^{-12ix}\)\\
    \end{tabular}
  \end{ruledtabular}
  \caption{Phases $\varphi_{\mu a}(m)$ of the numbers replacing the condensed bosons that satisfy the antisymmetry
    constraints for $m=4$. The phase $\varphi_{\mu a}$ for a given $\mu$ and $a$ can be read from the Table, in which
    $x:=\frac{2 \pi}{m}$.}
  \label{tab:phase}
\end{table}
Note that it is also possible to parametrize the bosons similarly to Eq.~\eqref{eq:RS-zsubi} and Eq.~\eqref{eq:RS-zsubj}
by using the generalized Gell-Mann matrices in Eq.~\eqref{eq:RS-constraint-su4-m2V2} that is adapted to $N$ and $m$.

Out of the $Nm$ bosons per site, $\frac{N}{n_{\mathbf{sub}}}m = m^{2}$ bosons are replaced by complex numbers satisfying
the antisymmetrization constraints. The Bogoliubov transformation can then be performed to diagonalize the quadratic
Hamiltonian, yielding $Nm-m^{2}=m(N-m)$ branches in the structural Brillouin zone. The resulting dispersive branches and
the number of these branches are identical to the results obtained with the multibosonic approach in
Sec.~\ref{sec:general-m} without the flat branches. Hence the same conclusion regarding the ordered color-moment can be
drawn, namely that the only Heisenberg system that can potentially retain the color-ordered ground-state is the \SU{4}
particles with $m=2$ on the square lattice.

\section{Discussion}
As seen in the previous considerations in the harmonic order, the flat branches we obtained with the multibosonic method
are related to the multipole moments requiring more than one ladder-operator action. Since the Heisenberg Hamiltonian
contains the bilinear term only, these transitions will thus result in localized branches in the quadratic order, and
they do not intervene in the reduction of the ordering. The reduction of the color order originates solely from the
fluctuations that come from the permitted decay channels that yield the dispersive branches.

The multiboson spin wave in \SU{2} spin-$S$ systems as in Ref.~\onlinecite{romhanyi_multiboson_2012} gives us an insight
to this method. When applied to a \SU{2} Heisenberg spin-$S$ systems to the harmonic order, $2S$ branches emerge in the
structural Brillouin zone from which one branch is dispersive and the rest are flat. The dispersive branch describes the
dipole moments of the spins on neighboring sites, i.e., one spin flip that results in the reduction of the fully
polarized state $S_{\text{max}}= \pm S$ by one quantum $\Delta S_{z} = \mp 1/2$. The flat branches correspond to the
higher-order transitions requiring more than one spin-flip, thus reducing the polarization by more than one quantum. It
turns out that the dispersive branch is identical to the dispersive relation one obtains with the conventional spin-wave
theory for spin $S$ (but in the fundamental irrep), and one obtains exactly one band. In contrast, higher \SU{N}
symmetries yield more than one dispersive branch due to the more intricate group structure. For $N>2$, there are more
possible ways to change a color, i.e., there are $N(N-1)/2$ pairs of ladder operators
($\hat{S}^{\mu}_{\nu},\ \mu \neq \nu$) whereas \SU{2} possesses only one pair of ladder-operators.

\begin{figure}[!t]
  \centerline{\includegraphics[width=0.99\linewidth]{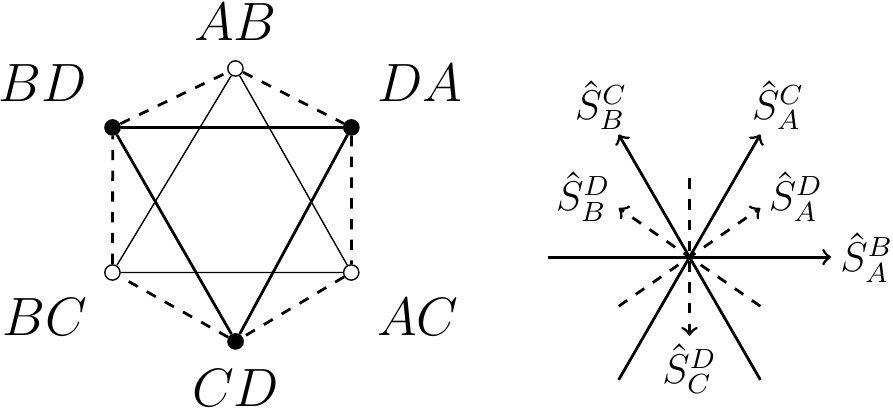} }
  \caption{\label{fig:su4-11}Left panel: weight diagram of \SU{4} in the antisymmetric $m=2$ irrep. The group \SU{4} being a group
    of rank 3, the states are characterized by three coordinates and the weight diagram is thus in 3D. The dots and the
    circles compose two different planes, the dots being on top of the circles. Right panel: the direction in which the
    six \SU{2} ladder-operators ($\hat{S}^{\mu}_{\nu}$, $\mu \neq \nu$) operate. The dotted lines have a non-zero
    component in the normal of the plane. One can attribute the labels $A$,$B$,$C$,$D$ to the states accordingly.}
\end{figure}

The accessible states by one color exchange can be schematically represented with the weight diagram of the
corresponding irrep, in which a state is associated to a point [an example of a weight diagram for the \SU{4} [0,1,0]
irrep is shown in Figure~\ref{fig:su4-11}]. For a given state, the states that can be reached by one color exchange are
the adjacent points on the weight diagram. The edges that connect points are in one of the $N(N-1)/2$ directions that
represent the action of the ladder operators $\hat{S}^{\mu}_{\nu},\ \mu \neq \nu$, and each direction is associated to
one specific color exchange. In our example, it can be seen that the a state in the irrep $\tiny{\yng(1,1)}$ of \SU{4}
has four adjacent points, thus showing the four states accessible by one color permutation. The action of the ladder
operators of \SU{N} can be depicted as the $N(N-1)/2$ directions in which the vertices between each point lie.

The Hamiltonian obtained through this method that describes the dynamics of these quantum fluctuations yields the same
dispersive branches as in the second method with Read and Sachdev bosons in Sec.~\ref{sec:altern-boson-repr}, although
the bosonic representations are different in both cases. The second approach has the advantage of containing exclusively
the physical branches at the quadratic order which contribute to the quantum fluctuations --- the flat multipolar
branches do not appear. Apart from these silent modes, they both give rise to the same results and yield identical
values of the ordered moment for each system we investigated.

According to the preceding analysis of the magnetization in Section~\ref{sec:multisun}, the color order persists in the
bipartite square lattice with two \SU{4} particles per site, but it is destroyed in the bipartite honeycomb lattice,
although the considered symmetry is identical (\SU{4}) in both bipartite lattices. This behavior is also observed in the
\SU{2} spin-$\frac{1}{2}$ AFM Heisenberg model. Comparing the values of the magnetization taken from
Ref.~\onlinecite{anderson_approximate_1952} and Ref.~\onlinecite{weihong_second-order_1991}, we observe that the
magnetization is smaller on the Néel honeycomb lattice than on the Néel square lattice. The smaller coordination number
$z$ of the honeycomb lattice leads to stronger quantum fluctuations, thus destroying the magnetic order. It is
worthwhile noting that the ratio of the reduction of the magnetic moment between the square lattice and the honeycomb
lattice $0.1966/0.2582 = 0.7614$ is the same as that of the reduction of the color moment of our models,
$0.7864/1.0328 = 0.7614$.

The tripartite triangular lattice with two \SU{6} particles per site also does not retain a finite color order. However,
the difference with the bipartite square lattice comes from the higher symmetry of \SU{6} in this case. As $N$ grows,
the number of decay channels of the quantum fluctuations becomes also larger. Hence, the quantum fluctuations are
stronger, and order is not favored as a consequence. As the study above involved the smallest non-trivial $m = N/k$
possible for each geometry, we expect that the only possible candidate for the color order with many particles per site
is the \SU{4} Heisenberg model on the bipartite square lattice. A pinning-field QMC study on this model has shown that
this model retains a finite magnetization of $m_i^{\text{alt}} \approx 0.24-0.26$ at their largest system size and
largest $U$,~\cite{wang_competing_2014} a value similar to our result in Eq.~\eqref{eq:magnetization-sq-2}. However, a
different QMC study shows results with no apparent broken lattice symmetry.~\cite{assaad_phase_2005} Hence, these
results call upon further investigation to settle the existence or non-existence of the magnetic order on this model.

\section{Conclusion}
We have applied the LFWT to systems with more than one particle per site described by fully antisymmetric \SU{N}
irreducible representations that are relevant to experiments with optical traps with more than one particle per site, first
in the spirit of the multiboson spin-wave theory and secondly using a different bosonic representation for antisymmetric
\SU{N} irreps. Both methods allow one to compute the ordered moment of the system and produce identical results. They
predict that the \SU{4} AFM Heisenberg model on the bipartite square lattice with two particles ($m=2$) retains a finite
long-range order even after including quantum fluctuations within the realm of the LFWT. The suggestion that this system
could be magnetically ordered allows one to potentially fill the corresponding point in the phase diagram of the \SU{N}
square lattice in Ref.~\onlinecite{hermele_topological_2011}. However, it is likely that the quantum fluctuations
destroy completely the color order for higher $N$ with $k=2$ as expected, due to the increase of quantum fluctuations
with increasing $N$. This is also true for the honeycomb lattice and the triangular lattice, where the ordered moment is
destroyed even for $m=2$, the smallest permissible $m$ assuming a two-sublattice order or a three-sublattice order,
respectively. The stronger quantum fluctuations in the bipartite honeycomb lattice compared to the bipartite square
lattice with the same \SU{4} symmetry are explained by the lower coordination number $z$ that reinforces quantum
fluctuations.

\begin{acknowledgments}
  We would like to thank Andrew Smerald and Mikl\'os Lajk\'o for useful discussions. This work has been supported by the Swiss National Science Foundation and by the Hungarian OTKA Grant No. K106047.
\end{acknowledgments}


\bibliography{bib}
\end{document}